\documentclass[a4paper,onecolumn]{IEEEtran}
\usepackage{hyperref}
\usepackage{filecontents} 		
\usepackage[nolist]{acronym}
\usepackage{amsmath,amssymb,amsfonts}
\usepackage{multirow}
\usepackage{mathrsfs}
\usepackage{graphicx}
\usepackage{verbatim}
\usepackage{tabularx}
\usepackage{subcaption}
\usepackage{caption}
\usepackage{textcomp}
\usepackage{smartdiagram}
\usepackage{adjustbox}
\usepackage{pgfplots}
\pgfplotsset{compat=1.18}
\usepackage{pdfpages}
\usepackage{parcolumns}
\usetikzlibrary{patterns}
\usepackage{float}
\usepackage{xcolor}
\usepackage{bm}
\usepackage[boxed,commentsnumbered, linesnumbered,ruled,vlined]{algorithm2e}
\usepackage[style=ieee, sorting=none]{biblatex}
\def\BibTeX{{\rm B\kern-.05em{\sc i\kern-.025em b}\kern-.08em
    T\kern-.1667em\lower.7ex\hbox{E}\kern-.125emX}}
\usepackage{epstopdf}
\usepackage{siunitx}
\usepackage{tikz}
\usepackage{circuitikz}
\usepackage{dirtytalk}
\usepackage{svg}
\usepackage{listings}
\usetikzlibrary{shapes.geometric, arrows, positioning}

\usepackage[left=2.2cm, right=2.2cm, top=3.5cm, bottom=2.2cm]{geometry}

\usepackage{titlesec}
\titlespacing{\subsubsection}{0pt}{10pt}{5pt}
\usepackage[skip=5pt plus1pt, indent=10pt]{parskip}

\usepackage{fancyhdr}
\lstdefinelanguage{Python}{morekeywords={async,await,cocotb.start_soon,dut,@cocotb.test(),CoverPoint,coverage_section,CoverCross,class,bins,rel,vname,assert,random,def,if,elif,return
	},morecomment=[l]{//}}
\graphicspath{{./figures/}}
\tikzstyle{startstop} = [rectangle, rounded corners, minimum width=3cm, minimum height=1cm,text centered, draw=black, fill=red!30]
\tikzstyle{io} = [trapezium, trapezium left angle=70, trapezium right angle=110, minimum width=3cm, minimum height=1cm, text centered, draw=black, fill=blue!30]
\tikzstyle{process} = [rectangle, minimum width=3cm, minimum height=1cm, text centered, text width=4cm, draw=black, fill=orange!30]
\tikzstyle{decision} = [diamond, minimum width=3cm, minimum height=1cm, text centered, draw=black, fill=green!30]
\tikzstyle{arrow} = [thick,->,>=stealth]

\begin{document}
	%\acrodefplural{SoCs}{System-on-Chips}
	%\acrodefplural{ICs}{Integrated Circuits}
	%\acrodefplural{IPs}{Intellectual Properties}
	%\acrodefplural{HDLs}{Hardware Description Languages}
	%\acrodefplural{HVLs}{Hardware Verification Language}
	\begin{acronym}[placeholder]		
		\acro{SoCs}{System-on-Chips}
		\acro{IPs}{Intellectual Properties}
		\acro{HDLs}{Hardware Description Languages}
		\acro{HVLs}{Hardware Verification Language}
		\acro{CRV}{Constrained Random Verification}
		\acro{BFMs}{Bus Functional Models}
		\acro{SoC}{System-on-Chip}
		\acro{GPI}{General Purpose Interface}
		\acro{MDV}{Metric Driven Verification}
		\acro{HDL}{Hardware Description Language}
		\acro{HVL}{Hardware Verification Language}
		\acro{ADC}[ADC]{Analog-to-Digital Converter}
		\acro{I2C}{Inter-Integrated Circuit}
		\acro{ALU}{Arithmetic Logic Unit}
		\acro{ASIC}{Application-Specific Integrated Circuit}
		\acro{AI}{Artificial Intelligence}
		\acro{IP}{Intellectual Property}
		\acro{Cocotb}{Coroutine based cosimulation testbench}
		\acro{TLM}{Transaction Level Modeling}
		\acro{UVM}{Universal Verification Methodology}
		\acro{DUT}{Design Under Test}
		\acro{BFM}{Bus Functional Model}
		\acro{EDA}{Electronic Design Automation}
		\acro{VPI}{Verilog Procedural Interface}
		\acro{VHPI}{VHDL Procedural Interface}
		\acro{PyUVM}{Python Universal Verification Methodology}
	\end{acronym}

\title{Effective Design Verification – Constrained Random with Python and Cocotb}

\author{
\IEEEauthorblockA{\vspace{4mm}Deepak Narayan Gadde,
	Infineon Technologies,
	Dresden, Germany
	(\textit{deepak.gadde@infineon.com})}
\IEEEauthorblockA{Suruchi Kumari, 
	Infineon Technologies,
	Dresden, Germany
	(\textit{suruchi.kumari@infineon.com})\\}
\IEEEauthorblockA{Aman Kumar, 
	Infineon Technologies,
	Dresden, Germany
	(\textit{aman.kumar@infineon.com})}
}

\maketitle

%this puts the logo on the title page as well
\thispagestyle{fancy}

%puts logo on all pages except title page
\lhead{\begin{picture}(-53,0) \put(-53,0){\includegraphics[height=2.2cm]{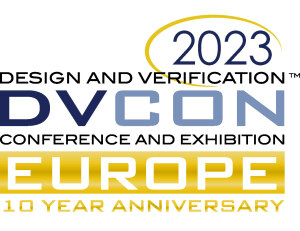}} \end{picture}}

\begin{abstract} 
\textbf{\emph{Abstract}\!
\textemdash Being the most widely used language across the world due to its simplicity and with 35 keywords (v3.7), Python attracts both hardware and software engineers. Python-based verification environment leverages open-source libraries such as \textit{cocotb} and \textit{cocotb-coverage} that enables interfacing the tesbenches with any available simulator and facilitating constrained randomization, coverage respectively. These libraries significantly ease the development of testbenches and have the potential to  reduce the setup cost. The goal of this paper is to assess the effectiveness of a Python-Cocotb verification setup with design \acs{IPs} and compare its features and performance metrics with the current de-facto hardware verification language i.e., SystemVerilog \cite{VerStudy_22}.}
\end{abstract}

\begin{IEEEkeywords}
\textbf{\emph{Keywords}\!
\textemdash \textit{cocotb; python; constrained random; functional verification; coverage; hardware verification language}}
\end{IEEEkeywords}

\section{Introduction}\label{sec:intro}
With the conclusion of Dennard scaling \cite{Dennard} and the deceleration of Moore’s law \cite{Moore}, the design of \ac{SoC} has become increasingly challenging. As the transistor size is shrinking at a remarkable rate, the total count of transistors  in a chip has increased exponentially over the years. This results to more functionality in the same die area, hence increased design complexity \cite{Hughes}. With such an increase in complexity of designs, the time required for verification experiences a significant upsurge. In comparison to directed tests, \ac{CRV} technique improves the productivity gain significantly \cite{Singh2004}. This methodology is crucial because it saves time in achieving coverage closure. The traditional method of using targeted tests to verify specific design elements grows exponentially with the number of inputs, hence there is a requirement for speeding up \ac{CRV} \cite{Hughes}.

There has been growing trend in the use of Python as high-level and general-purpose programming language. Renowned for its conciseness and formidable capabilities, it has emerged as a cornerstone for pioneering technologies, prominently encompassing artificial intelligence, automation, machine learning, \ac{GPI}, and networking \cite{gg}\cite{fund_py}. Additionally, the success of Python is due to (i) a simple and clean syntax, (ii) interpreted and dynamically typed (iii) object oriented (iv) huge ecosystem, (v) a rich standardized libraries that is conveniently available, and (vi) very good documentation and help support \cite{Cai}\cite{Fitzpatrick}.

\begin{figure}[H]
	\centering
	\begin{tikzpicture}
		\begin{axis}[
			xbar=0pt,
			xmin=5,
			width=11cm,
			height=8cm,
			xmax=1400,
			bar width=7pt,
			enlarge x limits={value=0.1,upper},
			legend pos={south east,font=\footnotesize},
			symbolic y coords={Python v3.7,C++, Java, C,C\#,Ruby, Smalltalk, Erlang, SV IEEE 1800-2009, SV IEEE 1800-2012},
			ytick=data,
			y tick label style={anchor=east,font=\footnotesize},
			nodes near coords,
			nodes near coords align={horizontal}
			]
			\addplot [fill=black!2,semithick,pattern = crosshatch dots,pattern color = black,font=\footnotesize] coordinates {(600,Python v3.7) (865,C++) (644,Java) (540,C) (511,C\#) (311,Ruby) (303,Smalltalk) (31,Erlang) (1285,SV IEEE 1800-2009) (1315,SV IEEE 1800-2012)};
			\addplot [fill=black!2,semithick,pattern = north east lines,pattern color = black,font=\footnotesize] coordinates {(35,Python v3.7) (83,C++) (50,Java) (32,C) (104,C\#) (42,Ruby) (6,Smalltalk) (28,Erlang) (221,SV IEEE 1800-2009) (248,SV IEEE 1800-2012)};
			\legend{spec\_pages,keywords}
		\end{axis}
	\end{tikzpicture}
	\caption[Language complexity with respect to number of specification pages and keywords]{Language complexity with respect to number of specification pages and keywords \cite{lang_complex}}
	\label{lang_complex}
\end{figure}
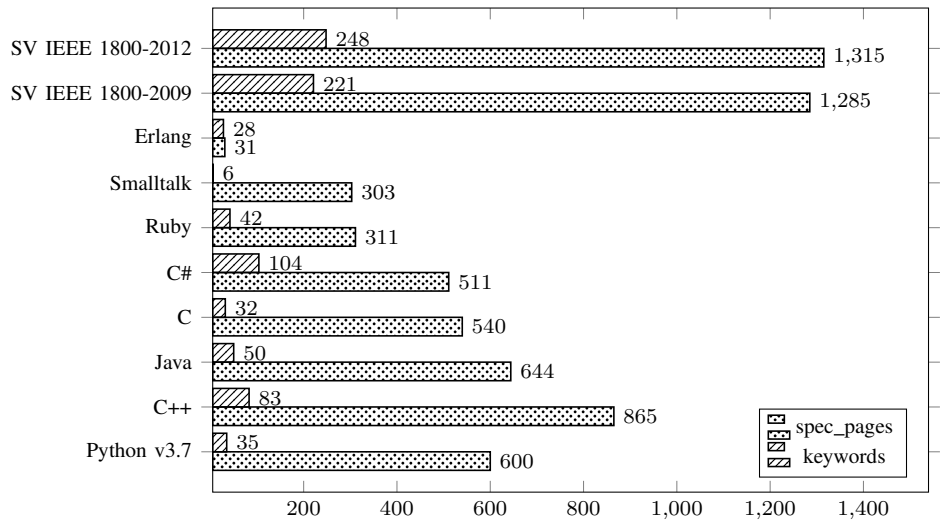

For the functional verification, \ac{HVL} Verilog was transitioned to SystemVerilog in order to incorporate numerous powerful programming features, particularly object-oriented programming with additional capabilities such as constrained random data generation and functional coverage \cite{Fitzpatrick}. SystemVerilog, which is also used as a language construct for industry-utilized verification methodologies like \ac{MDV} and \ac{CRV}, is a complex language with a steep learning curve especially for engineers who are new to hardware verification. It requires very good understanding of digital design concepts and a thorough grasp of the language syntax and constructs. Developing an efficient and robust testbench in SystemVerilog can be time-consuming \cite{Fitzpatrick}. Figure \ref{lang_complex} shows that \ac{HVL} i.e., SystemVerilog, is the most complicated language in comparison with other programming languages which has 1315 specification pages and 248 keywords as per IEEE 1800-2012. On the other hand, Python (v3.7) is a high-level programming language with only 35 keywords and 600 specification pages with 1750 full standard libraries \cite{lang_complex}. While SystemVerilog follows a set format, various \ac{EDA} tool vendors might incorporate the format differently within their simulation tools. Although it is widely used, there can be variations in the level of tool support across different vendors \cite{aldec}.

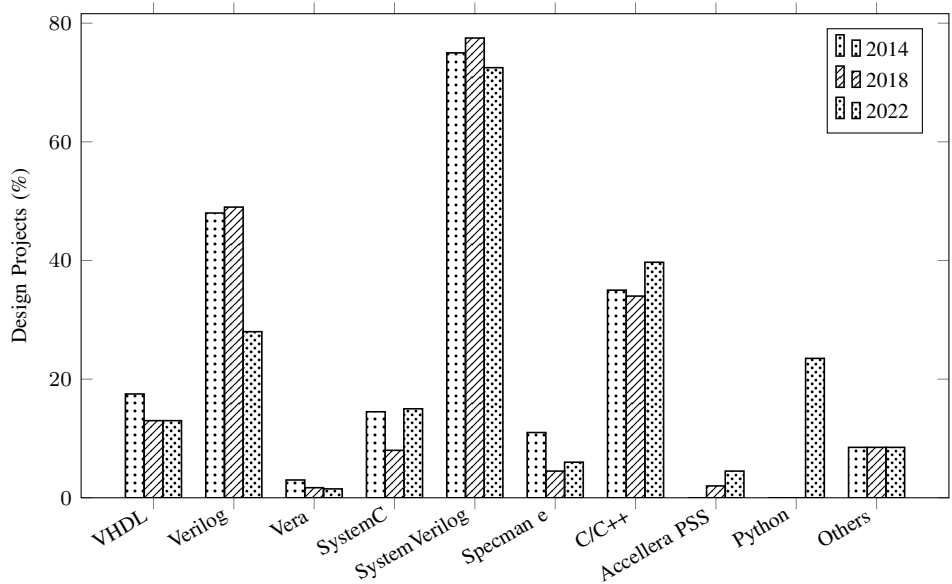
\begin{figure}[H]
	\centering
	\begin{tikzpicture}
		\begin{axis}[
			ybar=0pt,
			ymin=0,
			width=13cm,
			height=8cm,
			ymax=80,
			bar width=7pt,
			enlarge y limits={value=0.02,upper},
			legend pos={north east,font=\footnotesize},
			ylabel={Design Projects (\%)},
			symbolic x coords={VHDL, Verilog, Vera, SystemC, SystemVerilog, Specman e, C/C++, Accellera PSS, Python, Others},
			xtick=data,
			x tick label style={rotate=30,anchor=east,font=\footnotesize},
			label style={font=\footnotesize}
			]
			\addplot[fill=black!2,semithick,pattern = dots,pattern color = black,font=\footnotesize] coordinates {(VHDL,17.5) (Verilog,48) (Vera,3) (SystemC,14.5) (SystemVerilog,75) (Specman e,11) (C/C++,35) (Accellera PSS,0) (Python,0) (Others,8.5)};
			\addplot[fill=black!2,semithick,pattern = north east lines,pattern color = black,font=\footnotesize] coordinates {(VHDL,13) (Verilog,49) (Vera,1.7) (SystemC,8) (SystemVerilog,77.5) (Specman e,4.5) (C/C++,34) (Accellera PSS,2) (Python,0) (Others,8.5)};
			\addplot[fill=black!2,semithick,pattern = crosshatch dots,pattern color = black,font=\footnotesize] coordinates {(VHDL,13) (Verilog,28) (Vera,1.5) (SystemC,15) (SystemVerilog,72.5) (Specman e,6) (C/C++,39.7) (Accellera PSS,4.5) (Python,23.5) (Others,8.5)};
			
			\legend{2014,2018,2022}
		\end{axis}
	\end{tikzpicture}
	\caption[ASIC Verification Langauge Adoption]{ASIC verification language adoption \cite{VerStudy_22}} 
	\label{ver_lang_adoption}
\end{figure}

A recent study conducted by the Wilson research group in 2022 shows that the trend of using python as a \ac{HVL} for \ac{ASIC} development has been increased and can be seen in Figure \ref{ver_lang_adoption}. 

In this paper, the main focus is to evaluate the usage of a Python-based verification setup on design \acs{IPs} and compare it against the existing state-of-the-art \ac{HVL} for instance
SystemVerilog. The key highlights of the paper are listed below:
\begin{itemize}
	\item Comparison of SystemVerilog and Python as \acp{HVL}
	\item Exploration of Python libraries i.e., cocotb (framework enabling functional verification), cocotb-coverage (library allowing \ac{CRV} and functional coverage)
	\item Preparation of testbenches in Python and SystemVerilog for the design \acs{IPs}. i.e., 32-bit \ac{ALU}, \ac{I2C}, and 16-bit \ac{ADC}
	\item Analysis of results and empirical observations made during the testbench preparation for respective \acs{IPs}
	\item Comparison of performance metrics for both verification environments, simulated with various \ac{EDA} tools
\end{itemize}
The rest of the paper is organized as follows: Section \ref{sec:background} introduces functional verification along with various methodologies and presents related work. Section \ref{sec:implementation} explains the building blocks for the Python-\ac{Cocotb} verification implementation. Section \ref{sec:results} discusses the results and compares for SystemVerilog and Python-\ac{Cocotb} verification setups. Section \ref{sec:empirical} details the empirical observations made implementing the verification. Lastly, Section \ref{sec:conclusion} summarizes the paper and lists further scope of Python-based verification.

\section{Background}\label{sec:background}
The main goal of functional verification is to test the verification target, thereby guaranteeing accurate and comprehensive functionality \cite{Solr-EB000355927}. The present design verification stage utilizes a well- established technique for larger designs known as simulation- based verification. Such modern methodologies involve a largely automated procedure encompassing test generation, checking, and coverage collection, combined with instances of manual involvement \cite{8000621}. One of the methodology is \ac{CRV} where stimulus generation, scenarios can be generated in an automated fashion under the control of a set of rules, or constraints, specified by the user \cite{manish}. Another notable approach rooted in simulation is the \ac{UVM}, which employs \ac{TLM} for the creation of testbenches. It is a class library that makes it easy to write configurable and reusable code \cite{Singh2004}. All the technologies discussed uses SystemVerilog as their language construct. In contrast, this paper discusses simulation-based verification for 3 designs detailed in section \ref{sec:implementation}, using Python as \ac{HVL} and  \textit{cocotb}, \textit{cocotb-coverage} libraries.

The authors in paper \cite{ankitha} presents an explanation of the Python-based verification methodology along with a discussion on code coverage data obtained from verifying a design \ac{IP} using a Python testbench. A similar work is done in paper \cite{Cieplucha}, that extends \ac{Cocotb} to provide constrained randomization and functional coverage constructs. The paper also motivates to enable the implemented mechanisms to be adopted by verification engineers, taking advantage of Python syntax and object-oriented approach. Nevertheless, these works did not discuss feature, performance comparison with respect to the other verification methodologies. PyVSC is a library supported by Python that has the same functionality as cocotb-coverage library to support randomization and functional coverage \cite{Ballance}. The work claims that PyVSC will make it easier for SystemVerilog practitioners to reuse their knowledge of constraints and coverage in Python. The work \cite{naman} employs \ac{Cocotb} to model the Python testbench for a comparator designed in Verilog and an open-source simulator Icarus Verilog for simulating the testbench. Additionally, machine learning technique is implemented to optimize design verification. In the paper \cite{shibu}, the testbench is written in Python to develop a library (VeRLPy) for the verification of digital designs with reinforcement learning. 

The related works elaborated above mostly discusses creating Python testbench using \ac{Cocotb}, how SystemVerilog functional coverage constructs are supported by Python library like PyVSC. But none of them examines the comparison of the verification methodologies or language construct(s) used for the methodologies. In this work, we try to address the simulator compatibility, feature comparison between \ac{HVL}s, its performance in terms of run-time.

\section{Implementation}\label{sec:implementation}
The verification implementation with Python-\ac{Cocotb} involves three key components: \ac{DUT}, Testbench written in Python, and a Makefile. These building blocks play crucial roles in carrying out the verification process as discussed below.

\subsection{Design}
The \ac{DUT} in the Python-\ac{Cocotb} testbench can be designed in any \ac{HDL} i.e., SystemVerilog, Verilog, or VHDL. For this paper, we carefully selected three different designs for thorough verification. These choices were made based on important factors that enhance the significance of their verification. A short explanation to the design \acs{IPs} used in this work is as follows:

\subsubsection{ALU}
The 32-bit \ac{ALU}, a combinational design written in Verlilog, carries out a range of operations on the input signals and generates 32-bit output. It supports two arithmetic operations, namely addition and subtraction, and also provides functionality for six logical operations: NOT, AND, OR, XOR, NAND, and NOR. In Figure \ref{dut}, (a) depicts the block diagram of 32-bit \ac{ALU} where input buses \emph{a} and \emph{b}, control bus \emph{op}, and output bus \emph{r} are responsible for transmitting their corresponding signal data. Its computational capabilities finds applications in modern processors.

\subsubsection{I2C}
In general, reactive testbenches are particularly useful for protocol like \ac{I2C}, where the communication is event-driven and depends on the actions of both the master and the slave devices. For the verification process, the testbench created in Python-\ac{Cocotb} acts as Master for the design \ac{IP} coded in Verilog. In Figure \ref{dut}, (b) shows the block diagram of I2C master-slave configuration. \emph{A\textsubscript{0}}, \emph{A\textsubscript{1}}, \emph{A\textsubscript{2}}, and \emph{W\textsubscript{p}} are unidirectional input signals whereas \emph{SDA} and \emph{SCL} are bidirectional in/out pins. SDA transfers addresses and data during input and output operations; while SCL synchronizes the data exchange to and from the \ac{DUT}.

\subsubsection{ADC} 
\begin{figure}[H]
	\centering
	\includegraphics[width=0.9\linewidth]{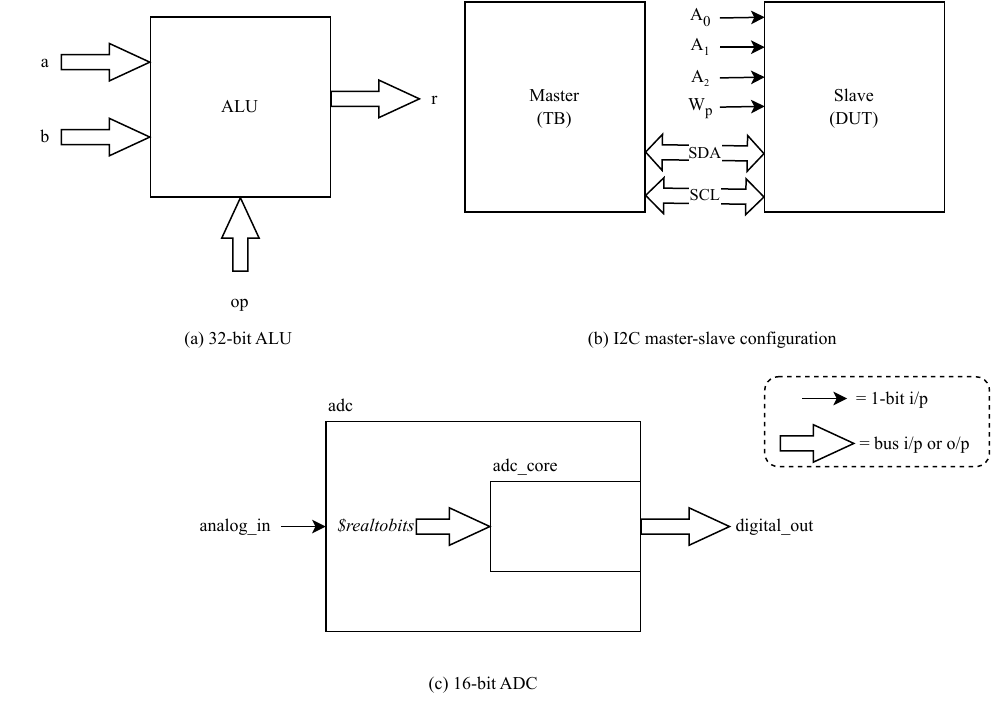} 
	\caption{Design(s) under test}
	\label{dut}
\end{figure}

It is an analog mixed signal design, implemented in SystemVerilog. Python-\ac{Cocotb} testbench is set up to evaluate \ac{Cocotb} with the design \ac{IP} performance. In Figure \ref{dut}, (c) shows that \emph{adc} is the top level, where real valued from -10V to 10V are given. \emph{\$realtobits} function has been used to convert \emph{analog\_in} to 64-bit analog input, which is then fed as input to \emph{adc\_core} and it produces the 16-bit digital output\emph{ digital\_out}.

\begin{comment}
	\begin{figure}[H]
		\centering
		\begin{subfigure}{0.4\textwidth}
			\centering
			\includegraphics[width=0.8\linewidth]{Images/dut_alu} 
			\caption{32-bit ALU}
			\label{dutalu}
		\end{subfigure}
		\vfill
		\begin{subfigure}{0.5\textwidth}
			\includegraphics[width=1.3\linewidth]{Images/dut_i2c}
			\caption{I2C master-slave configuration}
			\label{dutadc}
		\end{subfigure}
		\vfill
		%\hspace{40pt}
		\begin{subfigure}{0.5\textwidth}
			\includegraphics[width=1.3\linewidth]{Images/dut_Adc}
			\caption{16-bit ADC}
			\label{duti2c}
		\end{subfigure}
		
		\caption{Design(s) under test}
		\label{dut}
	\end{figure}
	
	\begin{figure}[H]
		\centering
		\includegraphics[width=0.35\linewidth]{Images/dut_alu} 
		\caption{32-bit ALU}
		\label{dutalu}
	\end{figure}
	\begin{figure}[H]
		\centering
		\includegraphics[width=0.6\linewidth]{Images/dut_i2c}
		\caption{I2C master-slave configuration}
		\label{duti2c}
	\end{figure}
	\begin{figure}[H]
		\centering
		\includegraphics[width=0.6\linewidth]{Images/dut_Adc}
		\caption{16-bit ADC}
		\label{dutadc}
	\end{figure}
\end{comment}

%\includepdf{Images/dut.pdf}
\subsection{Python Testbench}
\label{subsection:pytb}
The verification setup consists of a top testbench, where \ac{Cocotb} connects the Python testbench with the simulator. It also provides the Python library for creating synchronous logic. In this setup, the testbench uses constrained randomization of signals and bin definitions for analyzing functional coverage. These capabilities are supported by the \textit{cocotb-coverage} library. 

The basic structure of general Python-\ac{Cocotb} testbench is explained using \ac{ALU} testbench in Listing \ref{alutb}. Firstly, the important required modules are imported, for instance, \textit{cocotb}, and all the objects from \textit{cocotb-coverage} module. Then the tests specified in the testbench is automatically discovered by \ac{Cocotb} using \textit{cocotb.test()} decorator\footnote{Decorators use \textit{@} followed by the decorator name before a function or class declaration. When invoking the decorated function or class, the decorator runs first, and its output substitutes the original.} during simulation run. The clock is generated succeeded by signals being constrained and randomized. These signals are sent to \ac{DUT} and reference model. The coverage sample function is called and finally the outputs from \ac{DUT} and reference model are asserted. 

A single test specified in the \ac{ALU} testbench is simulated with various number of transactions i.e., \emph{20000}, \emph{40000}, and \emph{60000}. Similarly, a single test is present in the \ac{ADC} testbench which is run with \emph{210}, \emph{410}, and \emph{610} number of transactions.

\begin{figure}[H]
	\centering
	\includegraphics[width=\textwidth]{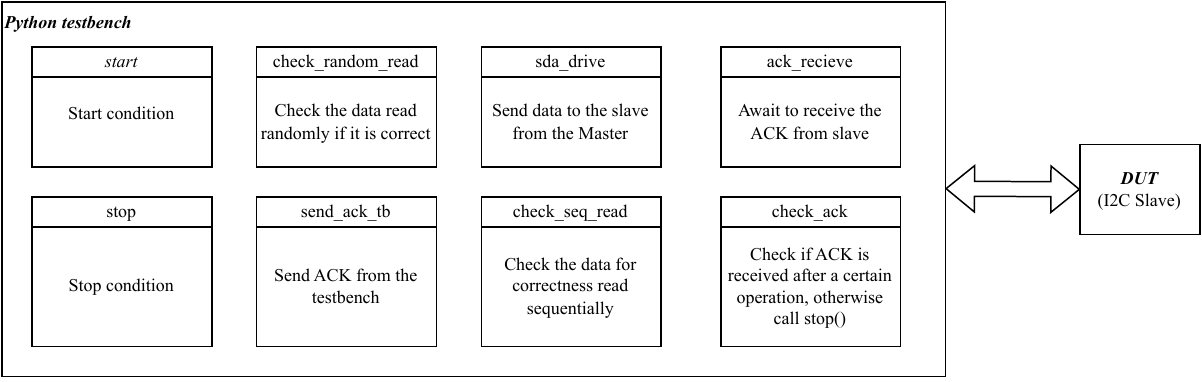} 
	\caption{BFM class based Python-Cocotb testbench e.g. I2C}
	\label{bfmi2c}
\end{figure}

A \ac{BFM} is a class that sends data using coroutine tasks. It contains coroutines\footnote{Coroutine is a decorated function that facilitates asynchronous execution and cooperative multitasking.} that manipulate signals to communicate to the \ac{DUT}. Hence, a class-based \ac{BFM} testbench is created for \ac{I2C} enhancing modularity. Figure \ref{bfmi2c} shows the \acs{BFMs} (coroutine tasks) included in the \ac{I2C} testbench (Master) and \ac{I2C} bus communicates betweeen testbench and the \ac{DUT}. The testbench includes 3 tests, namely, byte write followed by random read, page write succeeded by random read, and page write read sequentially.

\hspace{50pt}\begin{minipage}{.8\textwidth}
	\centering
	\begin{lstlisting}[language=Python, caption=Python-Cocotb testbench e.g. ALU,label={alutb},frame=tlrb]{Name}
		## Import required modules ##
		import cocotb
		..
		@cocotb.test() # Automatic test discovery
		async def alu_test(dut):
			""" Test coroutine starts with a handle to the the top level (dut)"""
			cocotb.start_soon(Clock(dut.clk,20,units="ns").start()) # Generate clock
			a = 0 # Initialize the signals
			..
			for i in range(1000): # No. of transactions specified in loop count
				ref_out = alu_ref(a,b,op) # Send the i/p signals to reference model and get o/p
				a = random. randint(-100, 100) # Randomize the signals
				..      
				dut.a.value = a      # Send the input signals to the DUT
				..
				sampling_function(a, b, op)  # Sample coverage
				assert ref_out == dut.r.value.integer # Assert outputs from reference model and DUT
		
	\end{lstlisting}
\end{minipage}
\subsection{Makefile}
A verification environment set-up needs a build option \textit{Makefile}. It contains information about the project, starting from \ac{EDA} tool to top level instantiation. Listing \ref{make} shows the Makefile for setting up the testbench environment for ALU. After the exceution of command in line 8, the libraries gets compiled and simulator starts.

%\includepdf{Images/python tb flowchart.pdf}

%\lstinputlisting[language=make, caption=Makefile (ALU), label={make}]{Listings/makefile}

\hfil\begin{minipage}{.8\textwidth}
	\centering
	\begin{lstlisting}[caption=Makefile e.g. ALU,label={make},frame=tlrb,firstnumber=1]{Name}
		SIM  ?= xcelium
		GUI  ?= 1
		TOPLEVEL_LANG ?= verilog 
		MODULE = alu_test
		TOPLEVEL = alu 
		VERILOG_SOURCES =  ../rtl/alu.v
		
		include $(cocotb)/makefiles/Makefile.sim
	\end{lstlisting}
\end{minipage}

\section{Results}\label{sec:results}
The three design \acs{IPs} i.e., \ac{ALU}, \ac{I2C} slave, and \ac{ADC} are verified in SystemVerilog-\ac{UVM} and Python-\ac{Cocotb} testbenches and investigated compatibility of \ac{Cocotb} with commercial simulators like Cadence Xcelium \cite{xcelium} and Siemens Questa \cite{questa}. Additionally, an open-source simulator Verilator \cite{verilator} is explored.
\subsection{Features Comparison}
While implementing the verification environments with SystemVerilog and Python as \ac{HVL}s, they are found to have setup advantages with Python over SystemVerilog like, setting up test environment with any simulator only requires to modify makefile variable \textit{SIM}. Table \ref{featurecom} shows other features compared for both verification implementations.

\begin{table}[H]
	\centering
	\caption{Features comparison for SystemVerilog and Python-\ac{Cocotb}}
	\setlength{\tabcolsep}{5pt}
	\resizebox{\textwidth}{!}{%
	\begin{tabular}{|l|l|l|l|} 
		\hline
		\textbf{Feature}                                                                                      & \textbf{SystemVerilog}                                           & \textbf{Python}                                                          & \textbf{Remarks}                                                                                                                                                                                             \\ 
		\hline
		\begin{tabular}[c]{@{}l@{}}\textbf{Declaration of }\\\textbf{data types}\end{tabular}                 & Static                                                     & \begin{tabular}[c]{@{}l@{}}Dynamic
		\end{tabular}                  & \begin{tabular}[c]{@{}l@{}}Python allows undeclared variables and perform \\any operation on them. Additionally, it has advanced data structures \\ e.g., tuple and dictionary, unlike SystemVerilog.\end{tabular}                \\ 
		\hline
		\begin{tabular}[c]{@{}l@{}}\textbf{Supported types }\\\textbf{of logic}\end{tabular}                  & \textit{0, 1, X, Z}                                              & \textit{X, Z, U, W}                                                      & Python-\ac{Cocotb} needs BinaryValue object for these logics                                                                                                                                                      \\ 
		\hline
		\begin{tabular}[c]{@{}l@{}}\textbf{Parameterization and }\\\textbf{size of the variable}\end{tabular} & Required                                                         & Not required                                                             & \begin{tabular}[c]{@{}l@{}}If size is not declared in SystemVerilog, data may be \\lost after an assignment to a different size than specified\end{tabular}                                                  \\ 
		\hline
		\textbf{Styles of control flow}                                                                       & \textit{begin, end}                                       & Proper indentation                                                       & \textit{elif }in Python replaces case in SystemVerilog/Verilog                                                                                                                                               \\ 
		\hline
		\textbf{Functions}                                                                                    & Not objects                                                      & Callable objects                                                         & \begin{tabular}[c]{@{}l@{}}Function in SystemVerilog are not objects and cannot be \\stored or passed directly as arguments\end{tabular}                                                                     \\ 
		\hline
		\textbf{Exceptions}                                                                                   & Not supported                                                    & Supported                                                                & In Python, exceptions are caught with \textit{try/except/finally} blocks                                                                                                                                     \\ 
		\hline
		\textbf{Libraries}                                                                                    & -                                                                & Large                                                                    & \textcolor[rgb]{0.114,0.114,0.114}{Create reference model for any complex design easily}                                                                                                                     \\ 
		\hline
		\textbf{Interpreted}                                                                                  & No                                                               & Yes                                                                      & \begin{tabular}[c]{@{}l@{}}\textcolor[rgb]{0.114,0.114,0.114}{It allows to restart the simulator without recompiling and }\\\textcolor[rgb]{0.114,0.114,0.114}{edit tests while it is running}\end{tabular}  \\ 
		\hline
		\textbf{Design Hierarchy}                                                                             & \begin{tabular}[c]{@{}l@{}}Includes top \\testbench\end{tabular} & \begin{tabular}[c]{@{}l@{}}Does not include \\top testbench\end{tabular} & \begin{tabular}[c]{@{}l@{}}It limits debugging capabilities, since tracing back signals \\in the testbench is not possible\end{tabular}                                                                      \\
		\hline
	\end{tabular}%
	}
	\label{featurecom}
\end{table}
\subsection{Performance Metrics}
Based on simulation results, certain performance metrics are defined. These metrics are analyzed and compared for both verification environments i.e., Python-\ac{Cocotb} and SystemVerilog-\ac{UVM} in this subsection.

\subsubsection{Design Hierarchy}
\begin{figure}[H]
		\centering
		\includegraphics[width=0.62\linewidth]{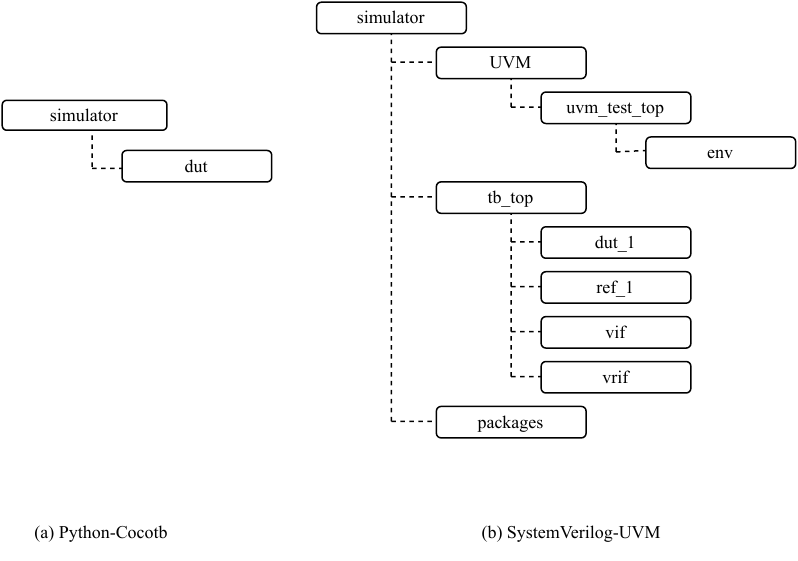} 
		\caption{Design Hierarchy}
		\label{deshierarchy}
	\end{figure}

	The hierarchy in the simulators design browser is different in SystemVerilog-\ac{UVM} and Python-\ac{Cocotb} testbenches. The SystemVerilog-\ac{UVM} testbench includes the top testbench whereas Python-\ac{Cocotb} starts with the \ac{DUT} or \textit{TOPLEVEL} defined in the build option, as explained in Figure \ref{deshierarchy}. This results in the limitation of Python-\ac{Cocotb} since it hinders the debugging capabilities of the testbench signals.
	\begin{comment}
		\begin{figure}[H]
			\centering
			
			\begin{subfigure}{0.25\textwidth}
				\includegraphics[width=1\linewidth]{Images/hierarchy_alu_py} 
				\caption{SystemVerilog/UVM}
				\label{despy}
			\end{subfigure}
			\begin{subfigure}{0.25\textwidth}
				\includegraphics[width=1\linewidth]{Images/hierarchy_alu}
				\caption{Python-\ac{Cocotb} }
				\label{dessv}
			\end{subfigure}
			
			\caption{Design Hierarchy (ALU)}
			\label{deshierarchy}
		\end{figure}
	\end{comment}

\subsubsection{Simulation run-time}
	The Python-\ac{Cocotb} and SystemVerilog testbenches are compared in terms of simulation run-time. They are simulated using various simulators such as Xcelium, Questa, and Verilator, with detailed testbench specifications provided in subsection \ref{subsection:pytb}. During the simulation, it became evident that Verilator cannot simulate \ac{I2C} and \ac{ADC} designs due to non-synthesizable nature of \ac{I2C} design used for this work whereas \ac{ADC} design required substantial modifications to verilate it. But \ac{ALU} design got simulated with Verilator, showing similar run-time performance as Xcelium. 
	
\begin{figure}[H]
		\centering
		\begin{subfigure}{0.495\textwidth}
			\includegraphics[width=1\linewidth]{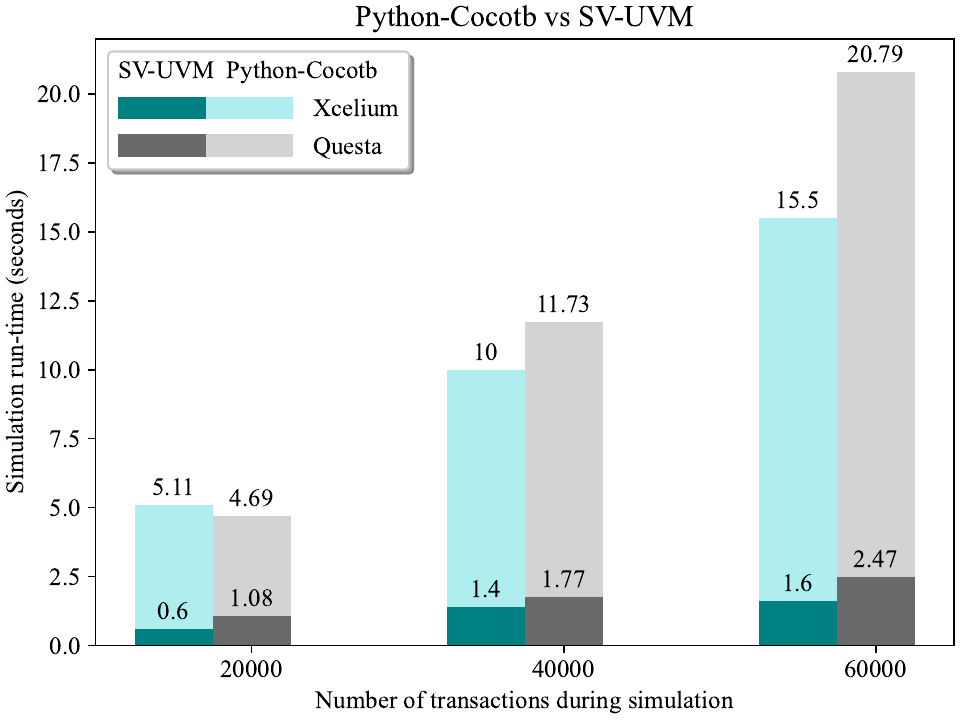} 
			\caption{ALU: A test is run separately for various transactions}
			\label{run_alu}
		\end{subfigure}
		\hfill
		\begin{subfigure}{0.495\textwidth}
			\includegraphics[width=1\linewidth]{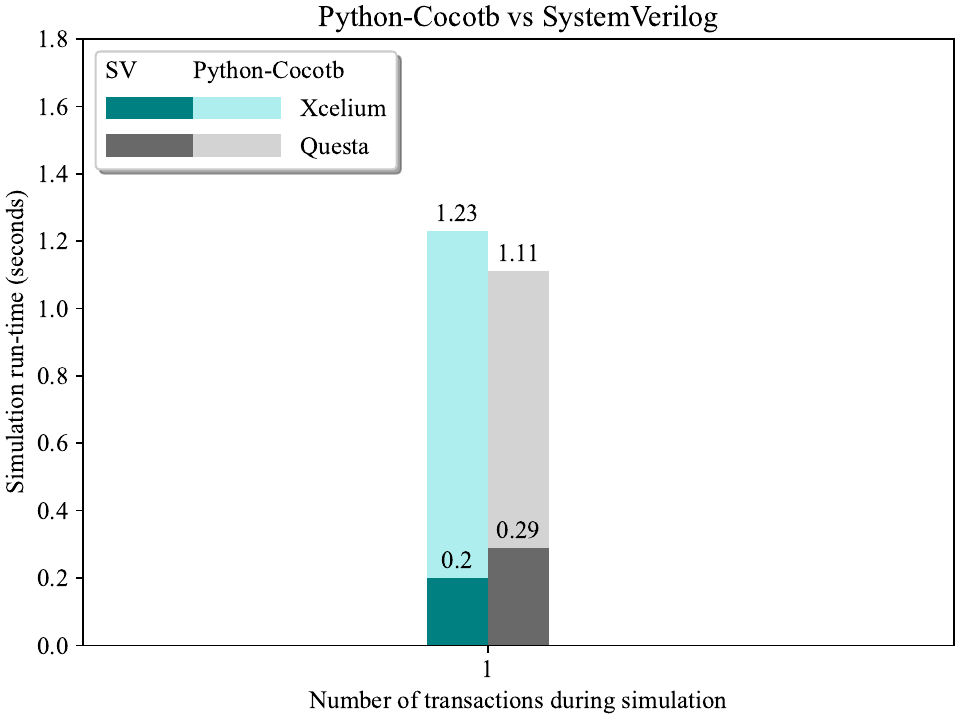}
			\caption{I2C: 3 tests are run in a single transaction}
			\label{run_i2c}
		\end{subfigure}
		%\medskip
		\begin{subfigure}{0.495\textwidth}
			\centering
			\vspace{10pt}
			\includegraphics[width=1\linewidth]{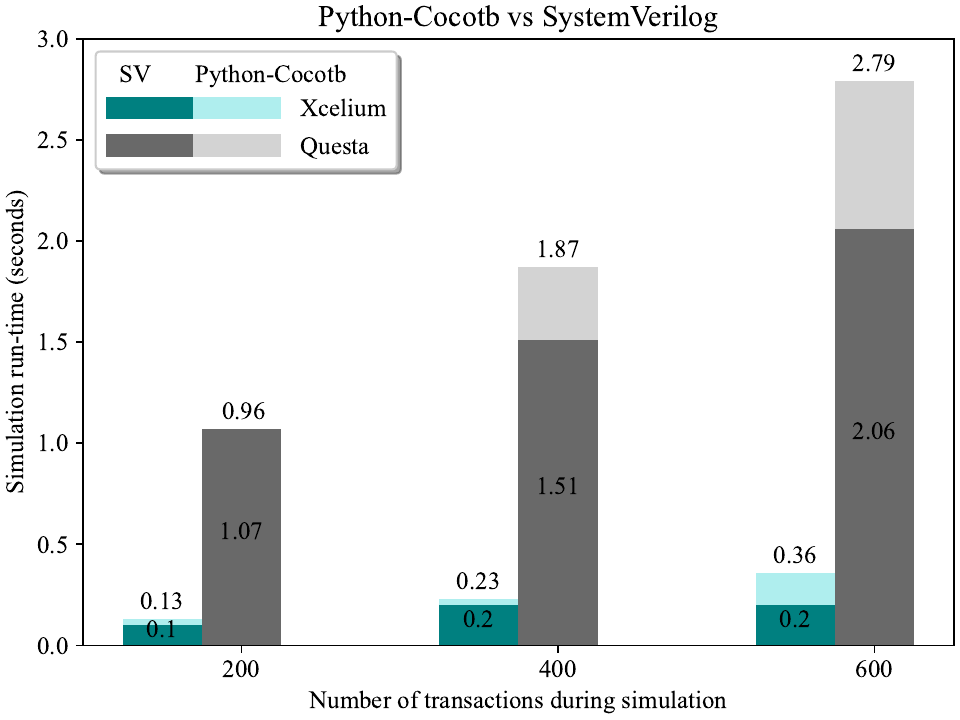}
			%\hspace{100pt}
			\caption{ADC: A test is run separately for various transactions}
			\label{run_adc}
		\end{subfigure}
		
		\caption{Simulation run-time comparison with different simulators}
		\label{run_time}
	\end{figure}

	Additionally, it is observed that the simulation run-time of Python-\ac{Cocotb} testbenches is slower compared to that of SystemVerilog or SV-\ac{UVM} testbenches, as demonstrated in Figure \ref{run_alu}, Figure \ref{run_i2c}, and Figure \ref{run_adc}. This difference is attributed to the reasons discussed. Generally, SystemVerilog employs simulation directives and commands to establish communication with the simulator. The close integration between SystemVerilog-\ac{UVM} and the simulator enhances simulation execution, resulting in relatively shorter run-time. Conversely, the interaction between Python testbenches and the simulator via \acs{VPI}/\acs{VHPI} is typically slower and less tightly integrated than the direct interaction between SystemVerilog-\ac{UVM} and the simulator. This overhead becomes more significant as the number of transactions increases, leading to longer simulation run-time for Python-\ac{Cocotb} testbenches.
 
\subsubsection{Coverage Analysis}
	Table \ref{covresult} details the bins defintion for all the cover items for design \acs{IP}, i.e., \ac{ALU}, \ac{I2C}, and \ac{ADC} along with the results. The coverage model defined was same in both SystemVerilog and Python testbenches and the results obtained were also similar in both cases. For \ac{ALU} and \ac{ADC}, the total coverage obtained in \SI{100}{\percent} for the bins defined whereas \ac{I2C} reached \SI{85.48}{\percent}. It was also verified that all the functional coverage features in SystemVerilog is available in \textit{cocotb-coverage} library, although they are syntactically different.

\begin{table}
	\centering
	\caption{Coverage results for SystemVerilog-UVM and Python-Cocotb testbenches}
	\setlength{\tabcolsep}{5pt}
	\resizebox{0.9\textwidth}{!}{%
	\begin{tabular}{|c|c|c|c|l|} 
		\hline
		\multicolumn{1}{|l|}{\textbf{Design}} & \multicolumn{1}{l|}{\textbf{Cover items}} & \multicolumn{1}{l|}{\textbf{Bins definition}}                                             & \begin{tabular}[c]{@{}c@{}}\textbf{Coverage}\\\textbf{\textit{(\%)}}\end{tabular} & \textbf{Description}                                                                                     \\ 
		\hline
		\multirow{7}{*}{\textbf{ALU }}        & a                                                  & \multirow{2}{*}{3 bins}                                                                            & \multirow{7}{*}{100}                                                                       & \multirow{2}{*}{Cover the range -2\textsuperscript{31} to 2\textsuperscript{31}-1}                                                                      \\ 
		\cline{2-2}
		& b                                                  &                                                                                                    &                                                                                            &                                                                                                                   \\ 
		\cline{2-3}\cline{5-5}
		& op                                                 & 1 bin                                                                                              &                                                                                            & Cover all operations ranging 0 to 7                                                                               \\ 
		\cline{2-3}\cline{5-5}
		& aXb                                                & \multirow{4}{*}{-}                                                                                 &                                                                                            & \multicolumn{1}{c|}{\multirow{4}{*}{-}}                                                                           \\ 
		\cline{2-2}
		& aXop                                               &                                                                                                    &                                                                                            & \multicolumn{1}{c|}{}                                                                                             \\ 
		\cline{2-2}
		& bXop                                               &                                                                                                    &                                                                                            & \multicolumn{1}{c|}{}                                                                                             \\ 
		\cline{2-2}
		& aXbXop                                             &                                                                                                    &                                                                                            & \multicolumn{1}{c|}{}                                                                                             \\ 
		\hline
		\multirow{9}{*}{\textbf{I2C }}        & c\_start                                           & \multirow{7}{*}{\begin{tabular}[c]{@{}c@{}} 2 bins; \textit{True, }\\\textit{False }\end{tabular}} & \multirow{6}{*}{50~}                                                                       & Cover start condition                                                                                             \\ 
		\cline{2-2}\cline{5-5}
		& c\_stop                                            &                                                                                                    &                                                                                            & Cover stop condition                                                                                              \\ 
		\cline{2-2}\cline{5-5}
		& c\_write                                           &                                                                                                    &                                                                                            & Cover write operation                                                                                             \\ 
		\cline{2-2}\cline{5-5}
		& c\_read                                            &                                                                                                    &                                                                                            & Cover read operation                                                                                              \\ 
		\cline{2-2}\cline{5-5}
		& c\_ack                                             &                                                                                                    &                                                                                            & Cover ACK recieved                                                                                                \\ 
		\cline{2-2}\cline{5-5}
		& c\_nack                                            &                                                                                                    &                                                                                            & Cover NACK condition                                                                                              \\ 
		\cline{2-2}\cline{4-5}
		& c\_repeated\_start                                 &                                                                                                    & 100                                                                                        & \begin{tabular}[c]{@{}l@{}}Cover repeated start, i.e., if write and\\read is done in the~same test.\end{tabular}  \\ 
		\cline{2-5}
		& c\_mem\_data                                       & 16 bins                                                                                            & 90.62                                                                                      & Cover for data write and read in memory are same                                                                  \\ 
		\cline{2-5}
		& c\_mem\_addr                                       & 32 bins                                                                                            & 100                                                                                        & \begin{tabular}[c]{@{}l@{}}Cover for write and read are same for\\memory address access\end{tabular}              \\ 
		\hline
		\textbf{ADC}                          & analog\_in\_tb                                     & 3 bins                                                                                             & 100                                                                                        & Cover the range -10V to 10V                                                                                       \\
		\hline
	\end{tabular}%
	}
	
	\label{covresult}
\end{table}

\begin{comment}
	\begin{figure}[H]
		\centering
		\includegraphics[width=0.7\linewidth]{Images/run_time_alu} 
		\caption{32-bit ALU - One test is run seperately for 20000, 40000, and 60000 transactions. Python-\ac{Cocotb} is compatible with Xcelium, Questa, and Verilator when used with ALU.}
		\label{run_alu}
	\end{figure}
	\begin{figure}[H]
		\centering
		\includegraphics[width=0.5\linewidth]{Images/run_time_i2c}
		\caption{I2C - 3 tests are written in the testbench where each test has a write operation followed by read operation such as Byte Write and Random Read, Page Write and Random Read, and Page Write and Sequential Read. The memory operations with data are randomized that is 1 byte long.}
		\label{run_i2c}
	\end{figure}
	\begin{figure}[H]
		\centering
		\includegraphics[width=0.5\linewidth]{Images/run_time_adc}
		\caption{ADC - The test is simulated thrice for 210, 410, and 610 transactions, in simulators Xcelium and Questa. The design requires significant modification to verilate for simulating with Verilator.}
		\label{run_adc}
	\end{figure}
\end{comment}

\section{Empirical Observations}\label{sec:empirical}
While implementing the verification enivironments for the design \acs{IPs} i.e., \ac{ALU}, \ac{I2C}, and \ac{ADC} with Python-\ac{Cocotb} and \textit{cocotb-coverage}, there are some observations made as listed below.
\begin{enumerate}
	\item ALU: While utilizing the Python-\ac{Cocotb} testbench (i) If the input signals \emph{a}, \emph{b}, and \emph{op} are not initialized, it gives \textit{Assertion Error} in the first clock cycle and simulation stops. Therefore, it is crucial to initialize the input signals before performing any operation. (ii) The bins definition for covering \ac{ALU} signals has to be specified explicitly. If auto bins are attempted to be created for this design \ac{IP}, it requires a significant amount of space for the number of bins to be generated and results in a memory error.
	\item I2C: SDA is open-drain terminal, so it has to be pulled up through a resistor. Python-\ac{Cocotb} lacks native support for pull-up/pull-down signals and tristate logic, a workaround is achieved by introducing an \ac{HDL} wrapper. In contrast, SystemVerilog, being a \ac{HDL}, provides built-in support for pulling up any signal, and does not need an additional wrapper. Figure \ref{i2cwrap} details the workaround to include tristate logic in Python testbench.
	
	\begin{figure}[H]
		\centering
		\includegraphics[width=0.5\linewidth]{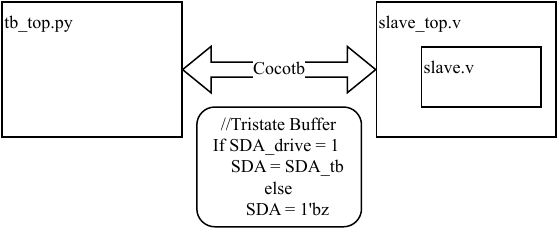} 
		\caption{HDL Wrapper to include tristate logic in I2C Python-Cocotb testbench}
		\label{i2cwrap}
	\end{figure}
	\item ADC: Analog simulation package that lets real number modelling in SystemVerilog, are not supported in the Python-\ac{Cocotb}. It gives VPI error (Communication error). To convert the real input \emph{analog\_in} to 64-bit digital input, the datatype is defined as \emph{real} type. This 16-bit signal is then sent to the design. Listing \ref{adc_sv} and \ref{adc_py} show that \emph{analog\_pack\_sv} module is imported in the \emph{adc} wrapper for SystemVerilog testbench whereas the input \emph{analog\_in} is declared as real datatype for Python testbench respectively.

\end{enumerate}

\hspace{50pt}\begin{minipage}{.8\textwidth}
	\begin{lstlisting}[caption=adc wrapper in SystemVerilog testbench, label=adc_sv, frame=tlrb,firstnumber=1]{Name}
		import analog_pack_sv : : * ;
		input analog_t analog_in ;
		assign analog_input = $realtobits(analog_in) ;
		
	\end{lstlisting}
\end{minipage}

\hspace{50pt}\begin{minipage}{.8\textwidth}
	\centering
	\begin{lstlisting}[caption=adc wrapper in Python-Cocotb testbench, label=adc_py,frame=tlrb,firstnumber=1]{Name}
		input real analog_in ;
		assign analog_input = $realtobits(analog_in) ;
		
	\end{lstlisting}
\end{minipage}

\section{Conclusion}\label{sec:conclusion}
In this paper, We analyzed the verification of three designs using Python-\ac{Cocotb} and SV-\ac{UVM} in three simulators. When running Python-\ac{Cocotb} testbenches, it is found that the simulation run-time increases with increase in transaction count for the multiple simulation run. This behavior is attributed to the communication between the testbench and simulators through \acs{VPI}/\acs{VHPI}, which involves longer interaction times with the simulators. Consequently, the overall simulation run-time is extended. Nevertheless, this interaction via \ac{GPI} provides necessary hooks to access and control the simulator’s internal data structures, signals, and events.

Concerning the \ac{CRV} and functional coverage in \ac{Cocotb}, the constrained randomization of the input signals and coverage constructs are derived from \textit{cocotb-coverage} library. The coverage analysis yields results similar to those obtained in SystemVerilog. 

Despite Python's interactive and user-friendly coding nature, the testbench is not included in the design hierarchy due to \ac{Cocotb}'s co-simulation approach. If the testbench can be integrated into the top of the hierarchy, it would greatly improve debugging capabilities by facilitating the tracing back of signals. Combining \ac{Cocotb} and Python libraries, and machine learning techniques holds promise for enhancing the verification process.

%\bibliographystyle{IEEEtran}
%\bibliography{IEEEabrv,paper}
\label{sec:bibliography}
\printbibliography[heading=bibintoc]%

\end{document}